
\magnification=1200
\hoffset=-.1in
\voffset=-.2in
\vsize=7.5in
\hsize=5.6in
\tolerance 10000

%
\def\one{  {\vcenter  {\vbox
              {\hrule height.4pt
               \hbox {\vrule width.4pt  height3pt
                      \kern3pt
                      \vrule width.4pt  height3pt }
               \hrule height.4pt}
                         }
                   }
           }

\def\two{  {\vcenter  {\vbox
              {\hrule height.4pt
               \hbox {\vrule width.4pt  height3pt
                      \kern3pt
                      \vrule width.4pt  height3pt
                      \kern3pt
                      \vrule width.4pt height3pt}
               \hrule height.4pt}
                         }
              }
           }
\def\oneone{ {\vcenter  {\vbox
              {\hrule height.4pt
               \hbox {\vrule width.4pt  height3pt
                      \kern3pt
                      \vrule width.4pt  height3pt }
               \hrule height.4pt
               \hbox {\vrule width.4pt  height3pt
                      \kern3pt
                      \vrule width.4pt  height3pt }
               \hrule height.4pt}
                         }
              }
           }
\baselineskip 12pt plus 1pt minus 1pt
\pageno=0
{MIT-CTP\#2337}\hfill{BRX-TH-358}

{BERC-94/101} \hfill{hep-th/9501099}
\bigskip
\centerline{\bf Universal Aspects of Two-Dimensional Yang-Mills Theory
at Large $N$}
\vskip 24pt
\centerline{Michael Crescimanno$^\dagger$
\footnote{*}{This
work is supported in part by funds
provided by the U.S. Department of Energy (D.O.E.) under
cooperative agreement \#DE-AC02-76ER03069 and by the
Division of Applied Mathematics of the U.S.
Department of Energy under contract \#DE-FG02-88ER25065.}
\footnote{$\ddagger$}{After Sept. '94 at {\it
Crescimanno@Berea.edu}, CPO 325, Physics Department, Berea
College,
Berea, KY, 40404.}}
\vskip 12pt
\centerline{\it Center for Theoretical Physics}
\centerline{\it Laboratory for Nuclear Science}
\centerline{\it and Department of Physics}
\centerline{\it Massachusetts Institute of Technology}
\centerline{\it Cambridge, Massachusetts\ \ 02139\ \ \ U.S.A.}
\vskip 24pt
\centerline{Howard J. Schnitzer
\footnote{$\dagger$}
{This work supported in part by funds
provided by the U.S. Department of Energy (D.O.E.) under grant
\#DE-FG02-92ER40706.}}
\vskip 12pt
\centerline{\it Department of Physics}
\centerline{\it Brandeis University}
\centerline{\it Waltham, MA ~~02254\ \ \ U.S.A.}
\vskip 1.0in

\centerline{\bf ABSTRACT}
\medskip

We show that the large $N$ partition functions and
Wilson loop observables of two-dimensional
Yang-Mills theories admit a universal functional form
irrespective of the gauge group.
We demonstrate that $U(N)$ QCD$_2$ undergoes a large $N$, third-order phase
transition on the projective plane at an area-coupling product of
$\pi^2/2$. We use this as a lemma to provide
a direct transcription of the partition functions
and phase portraits of
Yang-Mills theory from the $U(N)$ on ${\bf RP}^2$
at large $N$ to
the other classical Lie groups on $S^2$.
We compute the expectation value of the Wilson loops
in the fundamental representation
for $SO(N)$ and $Sp(N)$ on the two sphere.
Finally we compare the strong and weak coupling limit
of these expressions with those found elsewhere in the
literature.
\bigskip
\vfill
\eject
\baselineskip 24pt
\bigskip
\noindent{\bf SECTION {\it 1}.\quad Introduction}
\medskip
\nobreak
\noindent Recent work by many authors strongly suggests that pure
Yang-Mills
gauge theory in two-dimensions
is equivalent to a string theory$^{1-12}$. The interest in understanding
such a simple theory in such great detail is to shed some light
on the strong coupling limit of pure QCD in four dimensions. Of course,
the string picture of two-dimensional Yang-Mills
is interesting in it's own right as a testing ground
for non-perturbative analyses of a quantum field theory.

Studying two-dimensional pure Yang-Mills field theory
in the large $N$ limit is the starting point for making the
correspondence with string theory.
For example, as shown in Refs.[1,3,4],
a gauge theory based on $SU(N)$ at large $N$ splits into two copies
of a ``chiral'' theory that encapsulates the geometry of the string
maps.
The chiral theory associated to Yang-Mills theory on
a two-manifold $\cal M$ is a
sum over maps from a two-dimensional world sheets
(of arbitrary genus) to
the manifold $\cal M$.
This leads to a expansion in $1/N$ for the partition function and
observables that converges for all
area-coupling product for
target space $\cal M$ of genus one and greater.

The string expansion breaks down for small coupling when the
target space is the two-sphere, {\it i.e.} for Yang-Mills
on $S^2$.
Simply put, for the sphere
there are an infinite
number of string diagrams that contribute to, for example, the
leading term of free energy.
Thus, this term, as formulated in terms of string perturbation theory,
may fail to converge at a certain value for the coupling.
That there must indeed be such a difficulty
with the string expansion was first shown by
Douglas and Kazakov$^{13}$
(see also Ref.[14].)
In particular, they showed that
the free energy of QCD$_2$ on the sphere undergoes a third
order phase transition at a value of the coupling $A=A_c=\pi^2$.
Extensions of their work appear in Refs.[15-18] in which
this relevance of this phase transition was better understood
from the
point of view of large $N$ technique
and its application to the ``chiral'' sector of the
stringy picture of QCD$_2$ was carried out.
Finally, Boulatov$^{19}$,and Daul, et al. $^{20}$
have computed the expectation value of the
Wilson loop (in the fundamental representation)
on the sphere using the large $N$ saddle point found
in the work of Refs.$[13, 14]$.

In this note we present the large $N$ analysis
for QCD$_2$ on the
real projective plane and
compute the partition function on the
sphere (and on the projective plane)
for the classical Lie groups $SO(N)$ and $Sp(N)$
(by convention $Sp(N)$ has rank $N/2$). We also modify
the techniques of Refs.$[19,20]$ to study the expectation
value of the Wilson line in the fundamental representation
for these groups. We find that at large $N$
it is possible to adapt ``technology'' developed in Refs.[13-17]
to these other cases.
We were surprised by
the simplicity of the application of these
techniques and find an underlying universality that has
implications
for the observables and for the string
interpretation of the theory$^{22}$.

     In section {\it 2} we describe the large $N$ analysis of
$U(N)$ QCD$_2$ on the real projective plane.
These results serve as a lemma which allows
one to compute
the large $N$ phases of the gauge theories $Sp(N)$ and $SO(N)$
These conclusions are presented
in section {\it 3}.
In section {\it 4} we compute the Wilson loop in the fundamental
representation
on the two sphere for these gauge groups.
There we make contact with the known results in
strong and weak coupling and further describe the
universal analytic form of the Wilson loop and partition
function.

\bigskip
\noindent{\bf SECTION {\it 2}: The Phases of $U(N)$ QCD$_2$ on the
Projective Plane}
\medskip
\nobreak
\noindent As
was described in Refs.$[{13,14}]$,
the partition sum for $U(N)$ QCD$_2$
on the sphere is, for large $N$, dominated by a single
representation. In general the partition sum
for QCD$_2$ is
$$ Z_{\cal M} = \int [{\rm D}A] exp(-{{1}\over{4g^2}}\int_{\cal M} {\rm
Tr}(F\wedge^*F)).
\eqno(2.1)$$
and specializing to the case where ${\cal M} = S^2$, (2.1)
has been shown to be\footnote{*}{In what follows $A$ is
the dimensionless area-coupling product $A=g^2 N Area({\cal M})$.
This differs from the definition of ${\cal A}$, the
dimensionless area of Ref.[8,10].}
$$ Z_{S^2} = \sum\limits_{R} {dim}(R)^2 e^{-A C_2(R)/2N}
\eqno(2.2)$$
where $C_2(R)$  is the quadratic Casimir for the representation $R$
and $dim(R)$ its dimension. In the large $N$ limit
it is always possible to write (2.2) in terms of
a (quantum mechanical) path integral in a continuous
variable $h$ that is linearly related to the row lengths of the
Young diagrams for representations
$$ Z_{S^2} = \int \left[{\rm D}h\right]\  e^{-N^2 S_{eff}(h)}.
\eqno(2.3)$$
For a fixed gauge
group $G$
at large $N$ the partition sum takes the form of (2.3) with
$h$ and $S_{eff}$ depending on the group. For reference, note that for
the case of $U(N)$ considered previously in Refs.[13,14] one has
$$ {dim}(R) = \prod_{i<j}^{N} (1 - {{n_i-n_j}\over{i-j}})
; \qquad C_2(R) =Nr + \sum_{i=1}^{N} n_i(n_i-2i+1)
\eqno(2.4)$$
The $n_i$'s are the row lengths of the
Young diagram ($n_i \geq n_{i+1} ~~\forall i$)
and $r=\sum_i n_i$ is the total number of boxes in the Young diagram
for $R$.
For transcription to the large $N$ limit it is
convenient to define
$$ x = i/N;\qquad h(x) = -n(x)+x-1/2
\eqno(2.5)$$
and thus for the case of $U(N)$ one has
$$ S_{eff} = - \int_0^1 {\rm d}x \int_0^1 {\rm d}y~ \log(|h(x)-h(y)|)
+{{A}\over{2}}\int_0^1 {\rm d}x ~h(x)^2 + const.
\eqno(2.6)$$

The action (2.6) has a non-trivial saddle point. This is due to the
fact that although the exponential factors in
$C_2(R)$ (quadratic in $N$)
quickly become
very small, the $dim(R)$ factors grow precipitously, the
competition between these two factors being responsible for
the existence of a non-trivial large $N$ saddle point.
Thus, the positive powers of the $dim(R)$ factors
cause the sum to be dominated at large $N$ by a single
non-trivial representation $R^*$.
This is special to the two-sphere
(and also the projective plane), as the
partition sum of QCD$_2$ on closed surfaces of
higher genus will involve non-positive powers of
$dim(R)$.

Furthermore, as shown in Ref.[13], the
nature of the saddle point changes
dramatically between the small area ($A$) and large area limits.
This change is accompanied by additional terms in the
free energy $F = -{{1}\over{N^2}}\log(Z)$ that are non-analytic in
$A$.
{}From this point of view, the non-analyticity
results from
requiring $h(x)$ represent a
legal Young diagram (for which $n_i \ge n_{i+1} ~~\forall i$).
In Ref.[13], for $U(N)$ at large $N$, that
change occurs at an area of $A_c = \pi^2$ and that
it is associated with a third order phase transition.

The partition function for QCD$_2$ for closed non-orientable surfaces
is also known$^{6,8}$.
For example, the partition function on
the projective plane (${\bf RP}^2$) is (compare with (2.2))
$$ Z_{{\bf RP}^2} = \sum_{R = {\bar R}} dim(R) e^{-{{AC_2(R)}\over{2N}}}
\eqno(2.7)$$
Note that the sum in this case is only over self-conjugate
representations.
Again, in this sum there will be competition between the
dimension factor and the exponential suppression in the Casimirs,
and so we expect that there will also be a non-trivial
saddle point (i.e. a dominant representation).

At large $N$ this sum again has a simple form in terms of a path
integral. Following the example of discussed before, the partition function
of $U(N)$ Yang-Mills on ${\bf RP}^2$ may be written as a
path integral in the form of (2.3)
\vbox {$$
\eqalignno{
Z_{{\bf RP}^2} &= \int' \left[{\rm D}h\right]
{}~exp\biggl[-N^2\bigl(-{{1}\over{2}}\int_{0}^{1}
{\rm d}x \int_{0}^{1} {\rm d}y ~\log(|h(x)-h(y)|) \cr\nobreak
&+{{A}\over{2}}\int_{0}^{1}{\rm d}x ~h^2(x) +const'\bigl)\biggr]
&(2.8) \cr
}
$$
}
where $\int'$ means that
the path integration must necessarily
include only those $h$ that represent self-conjugate representations.
In (2.9) $ff.$ we impose the requirement that the
path integral (2.8) involve a sum only over self-conjugate
representations. Here we describe another, simpler way of
understanding the result we will find below. Note the similarity
between the actions
of (2.6) and (2.8). Scale out a factor of
2 from the exponent in (2.8) and replace $A$ there by $A/2$.
The resulting action is precisely half that of (2.6).
Of course,
an overall factor will not affect the nature of the
saddle point. However, the fact that the sum is only over
self-conjugate representations could affect this conclusion profoundly,
if it were the case that the dominant representation $R^*$ was
not self-conjugate. It is simple to see that this cannot be the case.
The contribution from a representation and its conjugate
to the sum must be the same: therefore, if there is to be a unique
saddle point it must be a self-conjugate representation.
Alternatively this may be thought of as a consequence of
the ${\bf Z}_N$ charge symmetry of the partition function.
The large $N$  self-conjugate saddle point
$R^*$
remains the dominating saddle point when
computing observables
(as long as the representations into which the observables
decompose have bounded ${\bf Z}_N$ charges.) By this reasoning
we expect that the phases of the theory on the projective
plane will be the same as the phases on the two-sphere, with the
only difference being that the phase transition
on ${\bf RP}^2$ between the
two phases occurs at $A = A_c = \pi^2/2$.

We now directly
implement the self-conjugacy condition in the
sum Eq.(2.7).
Although as described in the preceding
paragraph, we already know the answer,
but the direct method
we describe is useful for
studying QCD$_2$ based on the other classical
Lie groups. The requirement
that we sum only over self conjugate representations in $U(N)$
means that there is the additional constraint
$n_i = -n_{N-i+1}$ on the rows on the Young diagrams
that contribute to the sum. In large $N$ continuum variables this
implies that $h$ (see (2.5)) must satisfy
$$h(x) = -h(1-x).
\eqno(2.9)$$
A simple way to implement this condition is to define a new function
$g$ such that
$$
\eqalignno{
 h(x) &= \left\{
 \matrix{
  g(x)  &\hbox{}  0\ge x \ge 1/2,
 \hfill \cr
  -g(1-x) &\hbox{} 1/2 \ge x \ge 1,
 \hfill \cr
 } \right.
  & (2.10) \cr
}
$$
the function
$g$ being defined only on the interval $[0,1/2]$. Writing
the partition sum (2.7) in terms $g$ allows one to write
an `unrestricted' path integral (we apply the measure zero condition
$g(1/2) = 0$ later in selecting the solution.)
$$\eqalignno{
 Z_{{\bf RP}^2} &= \int \left[{\rm D}g\right]
exp\bigl[-N^2(\int_{0}^{1/2} {\rm d}x
\int_{0}^{1/2} {\rm d}y ~\log(|g^2(x)-g^2(y)|)
\cr
&+A\int_{0}^{1/2}{\rm d}x ~g^2(x) +const')\bigr]
&(2.11)\cr}
$$
Note the range of the integrations.
The saddle point that dominates this path integral is given
by the equation of motion
$$
P\int {\rm d}s {{u(s)}\over{g^2-s^2}} = P\int {\rm d}s'
{{u'(s')}\over{g'-s'}} = A/2
\eqno(2.12)$$
where we have defined $g'=g^2,\ s' = s^2$ and $u'(s'){\rm d}s' = u(s){\rm
d}s$.
This is one of the very simplest principle-value integral
equations to solve$^{21}$ for $u$. A general solution is
$$ u'(g') = {{b+{{A}\over{\pi}}(a-g')}\over{\sqrt{g'(2a-g')}}}
\eqno(2.13)$$
for some yet-to-be-determined constants $a$ and $b$.
Here $g' \in [0,2a]$.
Now, using the fact that $\int {\rm d}g u(g) = 1/2$, we can determine
the constant $b$
$$ u'(g') =
{{1}\over{2\pi}}\left[{{1+2A(a-g')}\over{\sqrt{g'(2a-g')}}}\right]
\eqno(2.14)$$
The solution $u$ must be bounded and this implies that
$a=1/2A$. Now
transforming $u'$ back to $u$ we have
$$ u(g) = {{A}\over{\pi}}{\sqrt{{{2}\over{A}}-g^2}}
\eqno(2.15)$$
where $g \in [-\sqrt{{{2}\over{A}}}, \sqrt{{{2}\over{A}}}]$.
This is the same solution found for the case
of $U(N)$ in Ref.[13], up to a scaling of $A$
by a factor of 2.

Since $h(x)$ must represent a legal Young diagram, we
must further require that $u\le 1$ everywhere.
For large areas ($A>\pi^2/2$) the solution
(2.15) will have regions for which $u>1$. Thus it is
necessary that we adopt an alternative Ansatz for solving the
integral equation (2.12) in that case. Just as in the $U(N)$ case, the
double cut Ansatz is applicable, and analysis reveals that
again the $U(N)$ form of $u(h)$ is precisely that found in the
large $A$ limit of the analysis of Douglas and Kazakov,
except for the overall factor of two in the area.

Furthermore, it is easy to check directly that the
free energy $F=\log(Z_{{\bf RP}^2})$ is simply that found
in Ref.[13] but scaled by an overall factor of 1/2 and
$A \rightarrow 2A$. Thus,
on ${{\bf RP}^2}$ at $A=A_c=\pi^2/2$ large $N$ QCD$_2$ (with gauge
group $U(N)$) has a third order phase
transition.

\bigskip
\noindent{\bf \quad SECTION {\it 3}: The Large $N$ Analysis of $Sp(N)$ and
$SO(N)$}
\medskip
\nobreak
\noindent It is possible to
take the large $N$ limit of gauge theory for any classical Lie  group
using
the methods discussed in the previous section. At first
sight it seems as though $Sp(N)$ and $SO(N)$ gauge theory
in the large $N$ limit in two dimensions would be quite different
than the case of $U(N)$ above.
The quadratic Casimir and the
dimensions of representations
differ quite significantly between the groups
$U(N)$, $Sp(N)$ and $SO(N)$, and the differences persist in the
large $N$ limit. Thus, somewhat surprisingly, we will show in this section
that their large $N$ phase structure (up to some simple factors)
is identical. The differences between the
various groups will first be manifest at non-leading order
in ${{1}\over{N}}$.

This surprising similarity between the different gauge models
suggests a very intriguing large $N$ universality in the
partition function and observables of these theories. We discuss this
universality in the context of the Wilson loop observable in QCD$_2$ in
section 4, and here study its antecedent in the partition function.
Recently, in collaboration with
S. G. Naculich, we have been able to  understand this universality
using the string picture$^{22}$.
{}From the group-theoretic point of view,
universality must thus
be a consequence of the ''unitarity'' (boundedness of the
eigenvalues), and not dependent on the
orthogonal or symplectic properties of the group elements.

To begin with, recall that the Casimirs for the
classical Lie groups are
$$ C_2(R) = f N\left[r-U(r)+{{T(R)}\over{N}}\right]
\eqno(3.1)$$
where
$$
\eqalignno{
f &= \left\{
           \matrix{
               1           & \hbox{ for } SU (N) \hbox{ and } SO (N),
                    \hfill \cr
               {{1}\over{2}} & \hbox{ for } Sp (N), \hfill \cr
           }     \right.  \cr  \noalign{\vskip0.5ex}
U (r) &= \left\{
           \matrix{
                   r^2 / N^2  &  \hbox{ for } SU (N), \hfill \cr
                   r / N      &  \hbox{ for } SO (N), \hfill \cr
                   -r / N     &  \hbox{ for } Sp (N),  \hfill \cr
           }     \right.
           &  (3.2)  \cr   \noalign{\vskip0.5ex}
T (R) &= \sum^{{\rm rank} G}_{i = 1}
           n_i (n_i + 1 - 2i)
           = \sum^{k_1}_{i = 1} n^2_i -\sum^{n_1}_{j = 1} k^2_j \> , \cr
}
$$
and where $n_i$ are the row lengths of the Young diagram ($k_i$ are
the column lengths)
associated to the representation $R$. Thus we have
$n_i \ge n_{i+1} \ge 0$ for all $1<i<rank(G)$. Note that the
solutions found by Douglas and Kazakov in Ref.[13] satisfy a
relaxed condition appropriate to $U(N)$, namely that the $n_i$
may be negative. However, by simply shifting the
independent variable one finds solutions for $SU(N)$ in which the
$n_i\ge 0 $. For the cases of
$Sp(N)$ and $SO(N)$ we will verify that our solutions
do indeed satisfy the constraint $n_i \ge 0$.

In terms of row lengths $^{23}$
$$ dim(R) = \prod_{i<j}^n {{(l_i^2 - l_j^2)}\over{(m_i^2 - m_j^2)}}
\qquad \qquad \hbox{ for } SO(2n)
\eqno(3.3)$$
where $l_i = n_i + n - i$ and $m_i = n-i$ while
$$ dim(R) = \prod_{i<j}^n {{(l_i^2-l_j^2)}\over{(m_i^2-m_j^2)}}
\prod_{i=1}^n {{l_i}\over{m_i}}
\qquad \qquad \hbox{ for } Sp(N) \hbox{ and } SO(2n+1)
\eqno(3.4)$$
with $l_i = n_i+n-i+1$ and $m_i = n-i+1$ for $Sp(N)$
where $N=2n$
while $l_i=n_i+n-i+1/2$, $m_i = n-i+1/2$ for $SO(2n+1)$.
Passing to the large $N$ limit, we find a universal form for
the $S_{eff}$ for the theories $Sp(N)$ and $SO(N)$
$$ S_{eff} =  -\int_{0}^{1/2} {\rm d}x
\int_{0}^{1/2} {\rm d}y~\log(|l^2(x)-l^2(y)|)
+{{A f}\over{2}}\int_{0}^{1/2}{\rm d}x~l^2(x) +const'
\eqno(3.5)$$
where $l(x)$ is the continuum (large $N$) limit of $l_i/N$.
This form is analogous to the equation (2.11) for the
large $N$ effective action of $U(N)$-QCD$_2$ on the
real projective plane ${\bf RP}^2$.

Of course, a crucial difference between the case of $U(N)$
and that of the groups $Sp(N)$ and $SO(N)$
is that,
for the later groups,
a solution related to an allowed
Young diagram must satisfy $n_i>0$ for all $i$.
However as we previously
emphasized on general grounds the solutions
discussed in Ref.[13] represent the $Q=0$ sector of the
$U(N)$ representation theory (see Ref.[15] for more details),
and furthermore must represent
self-conjugate representations.
Thus, the dominant
representation $R^*$ for all areas $A$ in the $U(N)$ theory
on $S^2$ or ${\bf RP}^2$ satisfies
$h(x) = -h(1-x)$, and so $h(1/2) = 0$.
Equivalently $g(x)$ of (2.10) is
positive over its entire range $0<x<1/2$.
Since $h(x)$ inherits monotonicity
from the $n_i$ we find indeed that the dominant
$U(N)$ representation has row lengths that are positive
between $[0,1/2]$ and negative beyond that point.

The transcription of (3.5) from the $U(N)$ case indicates
that we need only $1/2$ of the range of the
full $U(N)$ saddle point
distribution $h$,
which provides the
solution appropriate to the groups $Sp(N)$ and $SO(N)$
which have the constraint $n_i>0$.

It is also possible to compare the expressions for the
free energy in these various theories. In the light of the
above considerations it is perhaps not surprising that the
functional dependence of the free energy on the
area in these various theories are equal up to simple factors.
Indeed, using the equation of motion above, we find that
the derivative of the free-energy is
$$ {{{\rm d}F}\over{{\rm d}A}} ={{1}\over{2}} \int {\rm d}g~u(g)g^2.
\eqno(3.6)$$
Thus, we may immediately borrow the
results of the analysis of Ref.[13] to ascertain that
these other gauge theories on the sphere (or on the projective
plane) also have a strong and a weak coupling phase
separated by a third order phase transition. A summary of the
location of the critical points in the various theories
on $S^2$ is
$$ A_c = \pi^2/f
\eqno(3.7)$$
where $f$ is as defined in (3.1).
For ${\bf RP}^2$, all the critical areas are divided by two.

\bigskip
\noindent{\bf \quad SECTION {\it 4}: Wilson Loops on $S^2$ at Large $N$}
\medskip
\nobreak
\noindent In the preceding sections we have shown that there is a striking
large $N$ universality of the partition functions, phase portraits, and
free energies on the two-sphere between the classical Lie groups.
In an effort to extend this observation we present in this
section universal formulae for simple
Wilson loop observables in QCD$_2$.
Wilson loops for all $N$ in QCD$_2$ on the sphere
were first computed in Ref.[24].
We follow closely the notation, conventions
and arguments of Boulatov$^{19}$ and
Daul, et. al.$^{20}$ and refer the reader to relevant extensions
of these works in Ref.[25].

Consider a Wilson loop in the fundamental representation
$F$
dividing the two-sphere of (dimensionless) area
$A$ into regions of area $A_1$ and $A_2$
respectively.
Denote the expectation value of the
normalized Wilson line by $W_F$. In
two-dimensions the gauge-theoretic expression for
this Wilson loop
vacuum expectation value (VEV) is
$$ W_F = {{1}\over{dim(F)Z_{S^2}}}\sum_{R}\sum_{R'\in R\otimes F}
dim(R)dim(R')e^{-[{{{A_1}\over{2N}}C_2(R)+{{A_2}\over{2N}}C_2(R')}]}
\eqno(4.1)$$
where the second sum is over all the representations
$R'\in R\otimes F$.
and $Z_{S^2}$ is the partition function (2.1) for the theory on
a sphere of area $A = A_1+A_2$. It is now convenient to separate
the parts of the sum that involve terms proportional to
$e^{\pm N^2}$ raised to some (${\cal O}(1)$) power
from those
factors that do not have this $N$ dependence
as follows
$$ W_F = {{1}\over{dim(F)Z_{S^2}}}\sum_{R}
dim(R)^2 e^{-{{A}\over{2N}}C_2(R)}
\sum_{R'\in R \otimes F}
{{dim(R')}\over{dim(R)}}e^{-{{A_2}\over{2N}}[C_2(R')-C_2(R)]}.
\eqno(4.2)$$
The fundamental representation $F$ for the
classical Lie groups is denoted by a Young diagram
with a single box.
Thus, the Young diagram of $R'$ and $R$ differ only
by a single box, and so neither the ratio ${{dim(R')}\over{dim(R)}}$
nor the exponent $(C_2(R')-C_2(R))/N$ lead to terms proportional to
some power of $e^{\pm N^2}$.
The sum (4.2) over $R$ is again dominated by the same single `large $N$'
representation, $R^*$, which is the saddle point for the partition
function sum.

Refs.[19,20,24] give an account of the computation
of $W_F$ for the group $U(N)$ at large $N$.
We now show that,
surprisingly, the large $N$ VEV of
the Wilson loop $W_F$ has a
universal form independent of the
group. This is strongly reminiscent of the results
found in section {\it 3} for partition function.
It is a fully non-perturbative result.

Let us briefly review the results of Refs.[19,20].
Specializing to the case of $U(N)$ at large $N$, and adopting the
notation $\phi_i = i-n_i-N/2$, it is clear that $W_F$ is
$$ W_F = {{1}\over{N}}<\sum_j exp\left[\sum_i
\log\bigl(1-{{1}\over{\phi_j-\phi_i}}\bigr)+
\phi_i A_2\right]>
\eqno(4.3)$$
where the sum on the index $j$ represents
all the
rows of the Young diagram of $R^*$
that one can add a single box
to obtain an allowed diagram\footnote{*}{Equivalently,
one could average over $R'$, with one box {\it removed}
from $R^*$, and $A_1$ and $A_2$ interchanged, with the same
result$^{19,20}$. Note physically the VEV of $W_F$
on the sphere must be symmetric under
$A_1 \leftrightarrow A_2$.
These observations will be relevant to our discussion
of $Sp(N)$ and $SO(N)$.}.
As described in the literature,
the sum on $i$ in (4.3) is best approximated at large $N$ by sums
over two regions; those in which the box is added close to
the $j$'th row $|j-i|\leq \sqrt{N}$, and the other in which the
single box is put comparatively far from the $j$'th row,
$|j-i|\ge\sqrt{N}$. The first region contributes
an overall factor of
${{sin(\pi u)}\over{\pi u}}$, and the second, by
virtue of the equation of motion for $R^*$, combines with the
exponential term
that is already present, to yield,
$$ W_F = {{1}\over{N}}\sum_j exp\left[{{l_j \delta A}\over{2}}\right]
{{sin(\pi u)}\over{\pi u}}
\eqno(4.4)$$
where $\delta A = A_1-A_2$, and the sum is again over only those
rows $j$ of the Young diagram for $R^*$ that can legally admit
another box. Actually computing the sum over $j$ requires a little
care. This is due to the presence of rows to which one cannot add
even a single box. Taking account for such cases is
well described in Refs.[19,20] and it finally simply
changes an overall factor in the sum.
For $Sp(N)$ and $SO(N)$
we will have to address the very same problem.
In these cases it is similarly simple to prove that
accounting for these rows again simply results in
a change in the overall factor of the expression in the sum.

We now compute the VEV of a Wilson loop (in the fundamental
representation $F$)
for the groups $Sp(N)$ and $SO(N)$. Starting again with (4.2) and
specializing
to the case of $Sp(N)$ ($\eta = 1$, $f=1/2$;
for $SO(N)$ take ${\eta=-1}$ and $f= 1$)
we find
$$ W_F(Sp(N)) = {{1}\over{N}}<\sum_{R'} {{dim(R')}\over{dim(R^*)}}
exp\left[-{{A_1 {f}}\over{2N}}(C_2(R')-C_2(R^*))\right]>
\eqno(4.5)$$
In (4.5), $<>$ again denotes computing this average on the
dominant representation, $R^*$, of the $Sp(N)$ large $N$ partition function
on the sphere, while $\sum_{R'}$ is over all
representations that result by fusing $R^*$
and the fundamental representation $F$. For $Sp(N)$
this means the sum is over all rows of the
Young diagram of $R^*$ for which one can add one box
{\it and}
for which one can subtract one box, so as to
obtain a legal Young diagram.
The reason for this is that $Sp(N)$ has an anti-symmetric
invariant which allows (anti-symmetric)
contractions in the tensor product of the
fundamental representation with any
representation (a simple example in $Sp(N)$:
$\one \otimes \one = \two + \oneone + I$.)
As a consequence of using (3.3) and (3.4) for the dimensions
of representations, one has in the large $N$ limit
$$ W_F = {{1}\over{N}}< \sum_{j} \prod_{i\ne j} (1+{{1}\over{l_i-
l_j}})(1+{{1}\over{l_i+l_j}})
exp(-{{A_2 f}\over{2 N}} l_j)>
+ \{A_2 \leftrightarrow A_1\}
\eqno(4.6)$$
where the first term in (4.6) comes from adding one box and the second term
from removing one box. Both terms in (4.6) do indeed give the
same function (see the previous footnote)
However, as discussed in section {\it 3},
the constraint $n_i\ge 0$ for $Sp(N)$ means
that we have only $1/2$ of the range
of the full $U(N)$ saddle point distribution
$h(x)$. Because of the symmetry of the
saddle point of $U(N)$, taking $1/2$ the range
results in an overall factor of $1/2$ that
just cancels the overall factor of 2 in (4.6),
and so the Wilson loop for $Sp(N)$ is identical
in form to that of (4.4) with $A \rightarrow fA$.
Also, note that the question of nearby and distant
regions in the sum may be treated in exactly the
same way as that of $U(N)$.

The analysis of Wilson loops for $SO(N)$ is
identical. For $SO(N)$ fusions one recalls that
there is a symmetric (bilinear) invariant; one
must therefore again sum over all rows of
$R^*$ to which one can legally add one box
{\it and} subtract one box. As a result
the Wilson loop in the fundamental
representation (for all the
classical Lie groups) at large $N$ on $S^2$ can be
written in a universal way as
$$ W_F = {{1}\over{N}} \sum_j
{{sin(\pi u)}\over{\pi u}}exp\left[{{l_j f \delta A}\over{2}}\right]
\eqno(4.7)$$
where $l_j$ is defined for $U(N)$
above (4.3) and in (3.3), (3.4) for $SO(N)$, $Sp(N)$
respectively (see also (3.2))
replacing the sum by a continuous integral
results in the final universal result
$$ W_F (G;A_1,A_2) = \int {\rm d}\phi~
{{sin(\pi u)}\over{\pi}}exp\left[{{{f}\over{2}}(A_1-A_2)\phi}\right]
\eqno(4.8)$$
where  $\phi(u)$ is the continuous (large $N$) analogue
of $l_j$.
Note that in (4.8) the integration is over
the whole range of $\phi(u)$, due to the two identical terms in
(4.6) for $Sp(N)$ (as for $SO(N)$).

This proves the rather surprising result that there
is a universal large $N$ expression for the expectation
value of a Wilson line (in the fundamental
representation on the two-sphere)
that is independent of the gauge group.
This is expected for the small coupling phase for the
perturbative Wilson loop VEV but, we emphasize, here we
have shown a much stronger statement, namely that the
equality is a non-perturbative result
true for all coupling.
In the large area phase, $W_F$ is expressible
as an expansion in polynomials of $[A_i,e^{-A_i/2}]\quad i = 1,2$.
Now, in the light of
how different the dimensions and the Casimirs are between the
various classical Lie groups, we might expect that the actual
expansions in $[A_i, e^{-A_i/2}]\quad i = 1,2$
will differ between these groups. However,
because of the preceding analysis we know that
there is indeed a universal formula (4.8) which
evaluated in the large $A_i$ limit reads
\footnote{*}{To compare with the strong coupling
result of Boulatov$^{19}$ see appendix.}
$$\eqalignno{
 W_F &= e^{-A_2/2} + (-1-A_2+A_2^2/2)e^{-({2A_2+A_1})/2} \nobreak\cr
&+(1+A-{{A_1^2}\over{2}}+{{3A_2^2}\over{2}}+A_1A_2
-{{8A_2^3}\over{3}}
-{{A_1^2A_2}\over{2}}-{{5A_1A_2^2}\over{2}}+{{2A_1A_2^3}\over{3}}
+{{A_1^2A_2^2}\over{4}}+{{A_2^4}\over{2}})
e^{-(3A_2+2A_1)/2} \nobreak\cr
&+ ... + (A_2 \leftrightarrow A_1)
&(4.9) \cr}
$$
for both $SU(N)$ and $SO(N)$ (also for Sp(N) after the trivial scaling
$A_i \rightarrow fA_i$.) The fact that
these Wilson loop VEV's are the same
(regardless of the group) in the strong coupling regime is
not at all obvious from group theoretic considerations,
or the perturbative expansion of the theory
(since, at large $N$, we know we cannot follow
the usual perturbation expansion
into the strong coupling regime.)
It is simple to expand out the gauge
theory result about large $A_i$ to check (4.9).
The result is identical to the gauge-theoretic
normalized Wilson loop VEV formula at large $A_i$
(see for example section 5 of Ref.[10]).
{}From the gauge theoretic expression,
universality at large $A_i$
results from an intricate conspiracy between the
$dim(R)$ factors and the quadratic Casimirs.
It is {\it not} simply a fact about the leading
large $N$ part of the $dim(R) \sim d_R N^r/r!+$...
(in notation of Ref.[8])
but is a result that involves all the sub-leading
terms in the dimension formulae.
Of course,
it would be most satisfying to have a deeper understanding of this
universality. With S. Naculich, we have recently
completed
proof of this equivalence,
describing it in the string picture as an isomorphism
between string maps in the various theories. This proof of
large $N$ universality
will be presented elsewhere$^{22}$.

\goodbreak
\bigskip
\noindent{\bf \quad SECTION {\it 5}: Conclusion}
\medskip
\noindent We have shown the partition function and certain observables
on the two-sphere in QCD$_2$ at large $N$ are essentially
independent of
the underlying gauge group. This is not a trivial consequence of the
large $N$ limit, but emerges as
consequence of detailed analysis, and is true
for all values of the coupling-area product.
Such an equivalence is not obvious from the string point of view
but implies an equivalence between classes of maps
for any classical Lie group.

As described in forthcoming work$^{22}$, the string picture
of two-dimensional Yang-Mills gauge theories does provide
the most satisfying understanding of the universality
discussed in this present paper. This should be considered
a major success of the string picture of QCD$_2$ as it
gives insights not readily available otherwise.

\goodbreak
\centerline{\bf Acknowledgements}
\medskip
We thank S. G. Naculich, H. A. Riggs and W. Taylor
for comments.

\vfill
\eject

\vskip .3in
\centerline{\bf Appendix}
\medskip
\noindent This appendix is devoted to amending a few
of the
formulae in Ref.[19]. They regard the
computation of the
Wilson loop VEV at strong coupling.

In the strong coupling phase, much of the analysis
concerning
the large $N$
VEV of a Wilson loop in the fundamental representation involves
expansions of
elliptic functions. A reference on elliptic functions
that takes a very practical approach
(ideal for one checking the formulae in
Ref.[19]) is Lawden's book$^{26}$.
We will be brief in this appendix but
refer the interested reader
to that reference
for some of the details.

In Ref.[19] the second line of equation (50)
is in error.
Let $A = A_1+A_2$.
We find that in the
large coupling limit ($A \rightarrow \infty$ is the
$\tilde q = e^{-{{\pi a A}\over{4 K'}}}
\rightarrow 0$ limit, in the variables of Ref.[19])
$$ {{2K'}\over{\pi}}{{dn(2K'v, k')}\over{sn^2(2K'v, k')}}
= {{1}\over{\sin^2(\pi v)}} \left[1-4{\tilde q}+8{\tilde q}^2
+4{\tilde q}^2(\cos(2\pi v) + \sin^2(2\pi v)+\ldots \right]
\eqno(A.1)$$
All the other formulae in (50) of Boulatov$^{19}$ are
correct, but note the factor of $\pi$ difference between
the argument of the $\Theta$-functions in Ref.[19] versus
Ref.[26].

With this correction we modify (51) of Ref.[19]
for the VEV of the Wilson loop on the sphere to read
$$\eqalignno{
W_F &= -i\pi a \oint {{{\rm d}v}\over{2\pi i}}
{{exp{\left({{2\pi i K A_2}\over{K' A}} \cot(\pi v) + 2\pi i v\right)}}
\over{\sin^2(\pi v)}} \bigl[1-4{\tilde q}
+{{8\pi i K A_1}\over{K'A}}{\tilde q}^2\sin(2\pi v) \cr
&+8{\tilde q}^2+4{\tilde q}^2\left(\cos(2\pi v)+sin^2(2\pi v)\right)
+\ldots \bigr]
&(A.2)\cr}
$$
Note that the
saddle point implies that $4K = Aa$ and
now define
$\beta = 1+8{\tilde q}^2 \log {\tilde q}$. We then have
$$ {{a}\over{K'}} = {{1}\over{\pi}}\left(1
+8{\tilde q}^2 \log {\tilde q}+
\ldots\right)
= {{1}\over{\pi}}\beta + \ldots
\eqno(A.3)$$
Now, under the variable change $x = i\cot(\pi v)$ and also
inverting the relation for $a$ from equation (50) Boulatov
$$ a = {{1}\over{2}}\left(1+4{\tilde q}+4{\tilde q}^2
+8{\tilde q}^2\log{\tilde q}+...\right),
\eqno(A.4)$$
we may rewrite the Wilson loop VEV as
$$\eqalignno{
 W_F & ={{1}\over{2}}\oint {{{\rm d}x}\over{2\pi i}} \bigl({{1-x}\over{1+x}}
\bigr)e^{A_2\beta x/2}\bigl[1-4{\tilde q}^2
+4A_1{\tilde q}^2\bigl({{x}\over{1-x^2}}\bigr) \cr
& +8{\tilde q}^2\log{\tilde q}
- 4{\tilde q}^2\bigl({{1+x^2}\over{1-x^2}}+{{4x^2}\over{(1-x^2)^2}}
\bigr)+ \ldots\bigr]
&(A.5) \cr}
$$
For comparison see equation (51) of Ref.[19]. To the
order that we are working then we may replace all occurrences of
$\log{\tilde q}$ by $-A/4$. We must also be careful about keeping the
${\cal O}({\tilde q}^2)$ terms from the exponent
$e^{-A_2\beta/2} = e^{-A_2/2}(1+AA_2{\tilde q}^2)$
to the order that we are working. Putting all of this together
and computing the Wilson loop VEV
by summing the residues of the integral
(A.5), we find to this order the
result (4.9), as expected.

\vfill
\eject

\centerline{\bf REFERENCES}
\medskip
\item{1.} D.\ J.\  Gross, {\it Nucl. Phys.\/} {\bf B400} (1993) 161
\item{2.} J. Minahan, {\it Phys. Rev.} {\bf D47} (1993) 3430
\item{3.} D. Gross and W. Taylor, {\it Nucl. Phys.\/} {\bf B400} (1993) 181
\item{4.} D. Gross and W. Taylor, {\it Nucl. Phys.\/} {\bf B400} (1993) 395
\item{5.} D. Fine, {\it Commun. Math. Phys.} {\bf 134} (1990) 273;
{\bf 140} (1991) 321
\item{6.} E. Witten, {\it Commun. Math. Phys.} {\bf 141} (1991) 153
\item{7.} M. Blau and G. Thompson, {\it Int. Jour. Mod. Phys.} {\bf A7}
(1992) 3781
\item{8.} S. G. Naculich, H. A. Riggs, and H. J. Schnitzer, {\it Mod. Phys.
Lett.}
{\bf A8} (1993) 2223; {\it Phys. Lett.} {\bf B466} (1993) 466.
\item{9.} S. Ramgoolam, {\it Nucl. Phys.} {\bf B418} (1994) 30
\item{10.} S. G. Naculich, H. A. Riggs, and H. J. Schnitzer, 'The String
Calculation
of Wilson Loops in Two-Dimensional Yang-Mills Theory,' BRX-TH-355
(hep-th/9406100), S. G. Naculich and H. A. Riggs,
`The String Calcualtion of QCD Wilson Loops on Arbitrary Surfaces,'
BRX-TH-359,
(hep-th/9411143)
\item{11.} S. Cordes, G. Moore and S. Ramgoolam, `Large $N$ 2D Yang-Mills
Theory and Topological String Theory,' YCTP-P23-93, hep-th/9402107
\item{12.} P. Horava, `Topological Strings and QCD in Two Dimensions,'
preprint EFI-93-66, hep-th/9311156, November 1993
\item{13.} M.\ R.\ Douglas and V.\ A.\ Kazakov, {\it Phys. Lett.}
{\bf B319} (1993) 219
\item{14.} B. Rusakov, {\it Phys. Lett.} {\bf B303} (1993) 95
\item{15.} J. Minahan and A. Polychronakos, {\it Nucl. Phys.} {\bf B422}
(1994) 172
\item{16.} D.\ J.\ Gross and  A.\ Matytsin,
{\it Nucl. Phys. \/} {\bf B429} (1994) 50.
\item{17.} W. Taylor, 'Counting Strings and Phase Transitions
in 2D QCD', preprint MIT-CTP-2297, hep-th/9404175
\item{18.} M. Crescimanno and W. Taylor, 'Large $N$ Phases of Chiral
QCD$_2$', preprint MIT-CTP-2303
hep-th/9408115
\item{19.} D. Boulatov, {\it Mod. Phys. Lett.} {\bf A9} (1994) 365
\item{20.} J.\ M.\ Daul and V.\ A.\ Kazakov, {\it Phys. Lett.\/}
{\bf B335} (1994) 371
\item{21.} A. C. Pipkin,``A Course on Integral Equations,'' Berlin,
Springer Verlag,
1991
\item{22.} M. Crescimanno, S. G. Naculich and H. J. Schnitzer,
`The QCD$_2$ String and Large $N$ Universality', in preparation
\item{23.} W. Fulton and J. Harris, ``Representation Theory; A First
Course,''
New York, Springer Verlag, 1991
\item{24.} N. Bralic, {\it Phys. Rev.} {\bf D22} (1980) 3090
\item{25.} B. Rusakov, {\it Phys. Lett.\/} {\bf B329} (1994) 338
\item{26.} D. Lawden, ``Elliptic Functions and Applications,''
Berlin, Springer Verlag, 1989

\end